# On the threshold of ion track formation in CaF$_2$

M. Karlušić[1*], C. Ghica[2], R.F. Negrea[2], Z. Siketić[1], M. Jakšić[1], M. Schleberger[3], S. Fazinić[1]

[1] *Ruđer Bošković Institute, Bijenička cesta 54, 10000 Zagreb, Croatia*

[2] *National Institute of Materials Physics, Str. Atomistilor 105 bis, 077125 Magurele, Romania*

[3] *Fakultät für Physik and CENIDE, Universität Duisburg-Essen, D-47048 Duisburg, Germany*

[*] Corresponding author: marko.karlusic@irb.hr

There is ongoing debate regarding the mechanism of swift heavy ion track formation in CaF$_2$. The objective of this study is to shed light on this important topic using a range of complimentary experimental techniques. Evidence of the threshold for ion track formation being below 3 keV/nm is provided by both transmission electron microscopy and Rutherford backscattering spectroscopy in the channeling mode which has direct consequences for the validity of models describing the response of CaF$_2$ to swift heavy ion irradiation. Advances in the TEM and RBS/c analyses presented here pave the way for better understanding of the ion track formation.

## 1. Introduction

Material modification using swift heavy ions (SHIs) having a mass above 15 atomic mass units (amu), and a specific kinetic energy above 0.1 MeV/amu [FAL16] is an important contemporary research topic [CT09], [EA07], [PK08], [MCR13], [MCR11], [FA11], [NI09] with diverse applications like hadrontherapy [MT09], radiation waste storage [WJW98] and track etched membrane production [CT09]. Irradiation with SHI generates intense electronic excitation along its trajectory because at these kinetic energies (10-2000 MeV) the dominant channel for energy dissipation is via numerous collisions with electrons in the material. The density of deposited energy, usually expressed in terms of the SHI's electronic stopping power (keV/nm), is often quite sufficient to produce permanent damage along the ion trajectory (i.e. ion track) or activate various thermally driven processes enabling patterning or synthesis of nano-structured materials [MCR11], [MB09], [MB11], [IBR12].

Interactions of individual SHIs with crystalline materials can result in the formation of ion tracks in the bulk [PK08], [MCR13], [MT04], [MT12], [MB12], accompanied by surface features (nano-hillocks, nano-craters and similar) that can be found at the SHI impact site by means of atomic force microscopy (AFM) [EA07], [MK10], [FA11], [SA11], [OO15]. Despite their nanometric sizes, ion tracks in the bulk can be observed with several techniques. The most often used techniques are transmission electron microscopy (TEM) and Rutherford backscattering spectroscopy in the channelling mode (RBS/c) [MK10], [MT12], [AM94], [SOK04], [NK12], [MK15], [MT06]. TEM offers the possibility for direct observation of the individual ion tracks, while the RBS/c measures the ion track size indirectly by observing the fraction of disordered material in the crystal matrix, as a function of the applied SHI fluence. With some rare exceptions, the agreement in measured ion track radii between these two techniques is very good [MT06].

Due to the so-called velocity effect, slower ions (E/A < 2 MeV/amu) are much more efficient at ion track formation than the faster ones (E/A > 8 MeV/amu), even when both have the same electronic stopping power [AM93], [GS95], [MT04]. For this reason, medium sized accelerators can provide important complementary data at energies below 1 MeV/amu which are not easily accessible to the large accelerators. In particular, the threshold for ion track formation constitutes an important experimental quantity often needed for testing various ion-solid interaction models [MK12], [MK15]. Similar to ion tracks in the bulk, there is always a threshold for nanohillock formation on the surface. Above the threshold, nanohillocks typically grow in size with increasing ion energy. Similar values for the threshold for ion tracks in the bulk and for nanohillocks on the surface have often been observed [FA11], [NI09], thus enabling to characterize the threshold with different techniques.

Two most widely used thermal spike models, namely the inelastic thermal spike model (ITSM) and the analytical thermal spike model (ATSM) respectively, attribute the velocity effect to two completely different physical mechanisms. In the case of the ITSM, the velocity effect (i.e. reduction in track sizes at high SHI velocities) is related to the low density of the deposited energy due to the high energy of scattered electrons (primary or delta electrons) [MT12], [MT06]. This results in a monotonically increasing threshold for ion track formation with ion velocity, i.e. E/A. On the other hand, the ATSM attributes larger track sizes at low SHI velocities (below 2 MeV/amu) to a contribution from the Coulomb explosion [RLF65], [EMB02] and this is considered as an important difference between those two models [GS11], [GS13b].

## 1.1. Ion tracks in CaF$_2$: current status

At present, a hotly debated topic is the formation mechanism of ion tracks in CaF$_2$ and the closely related to this, the nature of the velocity effect (even its very existence) in this material [MT12b], [GS13], [MT13], [YYW14], [MK12], [GS00]. A still incomplete body of experimental data available for analysis is the main reason why this debate is ongoing. Despite extensive experimental work done previously [JJ98], [JJ98b], [NK05], [SAS07] where ion tracks in CaF$_2$ were investigated by TEM, only recently one experimental study using RBS/c was published [MT12b]. As noted by Szenes [GS13], this RBS/c data (along with XRD data from the same work) indicate the absence of the velocity effect in the 1-11 MeV/amu range. In response [MT13], TEM data was used as a proof for existence of the velocity effect in CaF$_2$ because very small ion tracks were found indicating a threshold for track formation around 20 keV/nm after GeV monoatomic irradiation, while a lower threshold around 10 keV/nm was found after MeV cluster ion irradiation. Not surprisingly, even the newest experimental results on AFM observation of SHI induced nanohillocks on CaF$_2$ surfaces were used in this discussion [YYW14].

The explanations given by the two models about the rather small track sizes observed in CaF$_2$ are radically different and at present seen as a test of their validity [MT12b], [GS13], [MT13]. According to the ITSM, the small track sizes observed by TEM are caused by strong ionic bonding that prevents amorphization within the ion track [Naguib]. Therefore in CaF$_2$, and probably other non-amorphizable materials, ion tracks consisting of strongly disordered material are a result of the quench of the boiling phase that occurs along the SHI trajectory [MT12b]. Melting, on the other hand, is seen as requirement for CaF$_2$ nanograin formation as observed by RBS/c and XRD [MT12b], as well as previous swelling studies [MB01], [MB02]. For this reason, the track radii measured by different methods are not always the same, and consequently the thresholds also vary. Based on the thermodynamic parameters for CaF$_2$, the amount of energy necessary to induce a molten phase is around 0.6 eV/atom [YYW14] and for the formation of the vapour phase around 1.7 eV/atom [MT12b] is required. For very low velocity SHI, in the ITSM this corresponds to stopping powers of 2.6 keV/nm [YYW14] and 7.5 keV/nm, respectively.

The problem of small ion track sizes in CaF$_2$ is solved within the ATSM in an altogether different way. Since large track sizes observed in other materials after irradiation with SHIs with low velocity are interpreted as contribution of a Coulomb explosion, small track sizes in CaF$_2$ are seen as evidence of its absence [GS13], [GS15]. Therefore, a pure thermal spike is thought to be at the origin of ion track formation in CaF$_2$, and in this later case all experimental data can be described using model parameters for the high velocity regime (i.e. $a_0$ = 4.5 nm, $g$ = 0.18), yielding a threshold at 9.5 keV/nm [GS00]. Recently, this threshold was revised down by Szenes to 7 keV/nm to accommodate newer RBS/c results, while the TEM data from refs. [NK05], [SAS07] were dismissed as erroneous [GS13]. Material melting remains as a necessary condition for ion track formation, while the boiling criterion is considered as unrealistic [GS13], although it is also possible to describe all TEM data available in the literature by applying boiling criterion within the ATSM and taking the velocity effect into account [MK12].

## 1.2. Ion tracks in CaF$_2$: motivation for the present work

The motivation for the present work is to resolve the dilemma outlined above by providing much needed additional experimental data. First, we observe that the TEM data of tracks in CaF$_2$ due to irradiation with monoatomic projectiles [NK05], [SAS07] show an unexpectedly uniform behaviour throughout the intermediate and high velocity range of the SHI (E/A = 2-11 MeV/amu). Similarly, RBS/c data [MT12b] show the same uniform behaviour in almost the same range of energies (E/A = 1-11 MeV/amu), as observed earlier by Szenes [GS13]. These observations led us to formulate the hypothesis that the velocity effect (as conceived in the ATSM) is shifted to lower values of E/A below 1 MeV/amu. Such a shift has been observed previously [GS99], but the magnitude of shift was much smaller.

To resolve the issues discussed above and to test our hypothesis, we have undertaken SHI irradiation of CaF$_2$ using monoatomic heavy ions with a specific kinetic energy E/A = 0.1 – 0.2 MeV/amu, followed by an investigation of the resulting ion tracks using TEM, RBS/c, AFM and time of flight elastic recoil detection analysis (ToF-ERDA). The SHI beams used in this study cover precisely the range of interest, i.e. electronic stopping powers between 3 – 7 keV/nm. Furthermore, in the present case the irradiation conditions match almost exactly the ones used in ref. [YYW14] because iodine and xenon ions at these energies are practically indistinguishable projectiles, hence the comparison between bulk and surface tracks in CaF$_2$ becomes possible. Thus, experimental data obtained here complements the existing data: in this energy range there is no RBS/c data, and available TEM data was obtained after MeV cluster ion irradiation, which could bear additional effects due to extreme values of deposited energy densities [SD07], [GS11b], [NAR00], [LTC03], [PS15] although opinions are divided on this topic [GS99], [MT12b]. This complementarity should enable us to establish the threshold for ion track formation in the bulk CaF$_2$, and to finally resolve the ardent controversy about the velocity effect in this material.

Finally, there is an open question about the atomic structure within the ion track. Based on previous investigations on the $CaF_2$ response to the electron beams [EJ83], the characteristic intermittent track structure observed by TEM is believed to consist of faceted anion voids [JJ98], [JJ98b], [NK05], [SAS07], [LTC03]. These voids can be viewed as calcium inclusions that should be easily formed due to small misfit with the fcc structure of the Ca sublattice and $CaF_2$. Then, the much lower density of the calcium with respect to $CaF_2$ would give rise to contrast as seen in TEM [JJ98b]. However, the fate of the fluorine gas liberated from within the ion track remains unknown: whether it remains trapped within $CaF_2$ crystal but outside the ion tracks [LTC03] or whether it is released by diffusion from the ion track [SAS07]. To address this particular question, both ToF-ERDA and scanning transmission electron microscopy – electron energy loss spectroscopy (STEM-EELS) have been performed in our current study.

## 2. Experimental details

Single crystalline $CaF_2$ (111) samples of 7x7 mm$^2$ have been prepared by cleaving from single crystal piece (Crystec) before the irradiation. Additionally, several TEM grids containing $CaF_2$ crystal grains have been prepared by crushing $CaF_2$ into a mortar followed by dispersion into ethanol and dripping the dispersion onto a TEM grid provided with a carbon membrane.

All SHI irradiations have been performed at the Ruđer Bošković Institute (RBI) using 6 MV EN Van de Graaff accelerator. Iodine ions with energies of 10 MeV, 15 MeV and 23 MeV have been used in both normal and grazing incidence geometry. Single crystal samples irradiated at normal incidence have been oriented with a small tilt angle (6 degrees) with respect to the surface normal, in order to avoid possible SHI channeling. The SHI beam has been scanned to ensure homogenous irradiation and the fluence has been measured by observing the ion flux in the Faraday cup before and after the exposure. For longer exposures, the irradiation has been interrupted a few times to ensure that the ion flux was stable. A full list of all irradiation parameters is given in table 1.

| SHI | $dE_e/dx$ (keV/nm) | $dE_n/dx$ (keV/nm) | R (μm) |
|---|---|---|---|
| 10 MeV I$^{5+}$ | 2.93 | 0.63 | 3.29 |
| 15 MeV I$^{6+}$ | 4.27 | 0.47 | 4.47 |
| 23 MeV I$^{6+}$ | 6.26 | 0.35 | 5.87 |
| 2 MeV Li$^{2+}$ | 0.72 | 0.001 | 3.82 |
| 1 MeV H$^{+}$ | 0.005 | 0.00004 | 11.99 |

**Table 1.** Irradiation parameters used in this work calculated using the SRIM code [srim]: electronic stopping power $dE_e/dx$, nuclear stopping power $dE_n/dx$ and projected ion range R.

Ion tracks in the bulk have been characterized using RBS/c and TEM. The samples have been analyzed by TEM before and after irradiation with SHI using a JEOL 200CX electron microscope operated at 200 kV acceleration voltage (NIMP) at room temperature. In order to further understand the processes induced by heavy ion irradiation but also by electron irradiation during the TEM investigation, the sample irradiated at 23 MeV has been analyzed by STEM-EELS technique. We have been using the JEM ARM 200F electron microscope corrected for spherical aberration in the STEM mode. The microscope has been operated at 200 kV, while the sample has been maintained at room temperature during observation.

For the RBS/c analysis, a 2 MeV Li$^{2+}$ and 1 MeV H$^+$ beam delivered by the 1 MV Tandetron accelerator has been used (RBI). The samples have been mounted on a goniometer and angular scan maps (tilt, azimuth) have been performed for target alignment. The beam spot size has been 1 mm in diameter and the ion current has been kept at about 1 to 2 nA. To detect the backscattered ions, a silicon surface barrier (SSB) detector with a thickness of 300 μm and bearing 3 mm slits has been positioned at 160° with respect to the probing beam direction.

Ion tracks on the surface have been analyzed using tapping mode AFM and *in situ* ToF-ERDA. The AFM measurements have been performed under ambient conditions using a Dimension 3100 AFM and Nanosensors NCHR cantilevers at Universität Duisburg-Essen (UDE). Images have been analyzed using the WSxM code [wsxm]. The ToF-ERDA measurements (RBI) have been performed using a 23 MeV I beam at 1° grazing incidence angle with respect to the sample surface. The ToF-ERDA spectrometer [tof1], [tof2] has been positioned at an angle of 37.5° towards the beam direction (RBI). All data have been collected in the 'list mode' and offline replay/analysis with sections has been performed using the Potku software package [potku].

## 3. Experimental results

### 3.1. TEM results

*Before irradiation*

At low TEM magnification the $CaF_2$ grains proved to be beam resistant. Working for long minutes in such low-beam illumination conditions did not induce any visible modifications at the morphological or structural level. However, after a few tens of seconds of e-beam irradiation under a focused e-beam, $CaF_2$ started to decompose, followed by Ca oxidation. One can observe in the following pictures that after exposing the crystal grain to a focused electron beam, it starts to modify both morphologically (as one can see in the TEM images) and structurally (see the selected area electron diffraction (SAED) patterns at the beginning of the TEM observation and after ca. 1 min. of observation under the focused beam). Diffuse diffraction rings appear in the SAED patterns along with the $CaF_2$ spots. The diffraction rings indicate the in-situ formation of CaO (cubic phase). In order to avoid the rapid beam-induced structural modifications, for the track studies we have chosen to work at low magnification using the diffraction contrast.

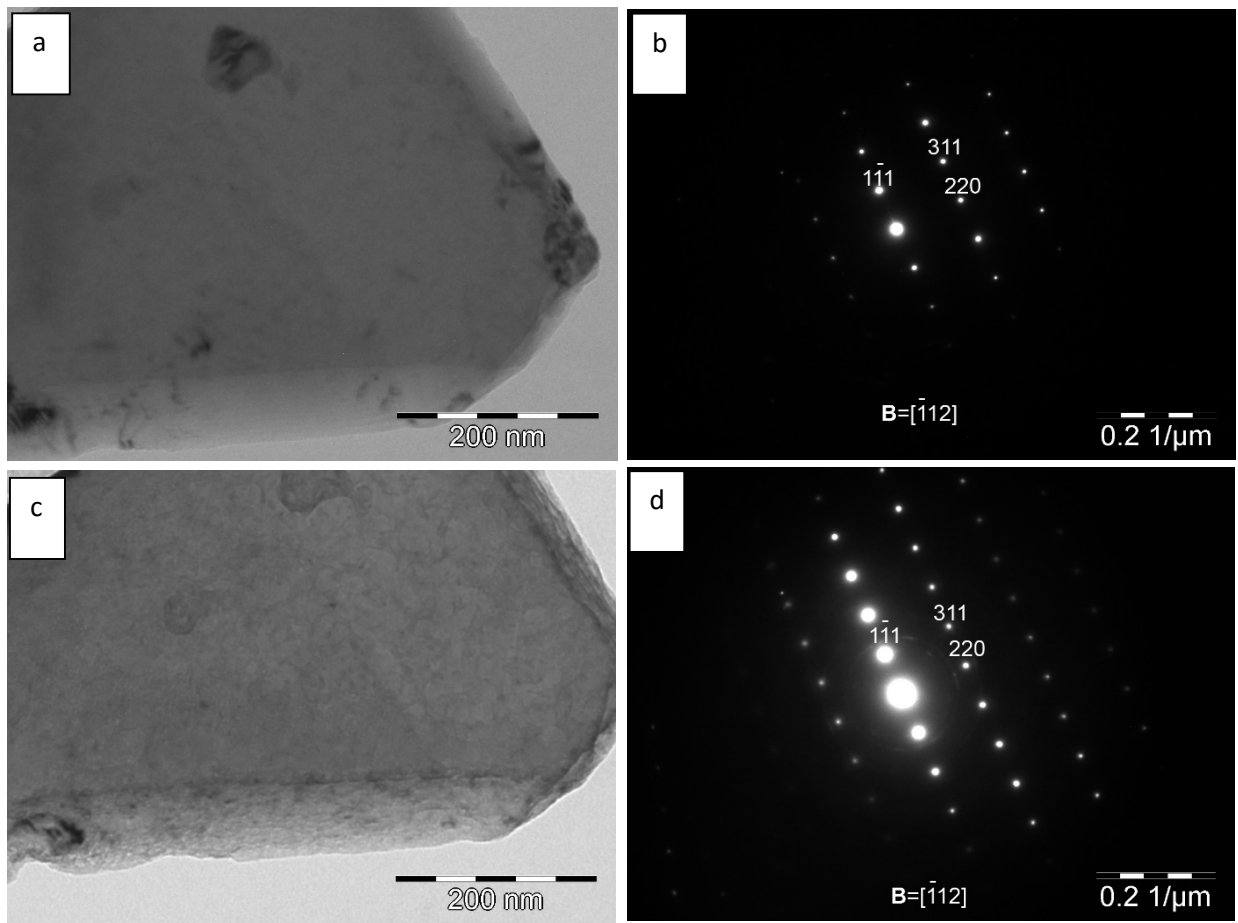

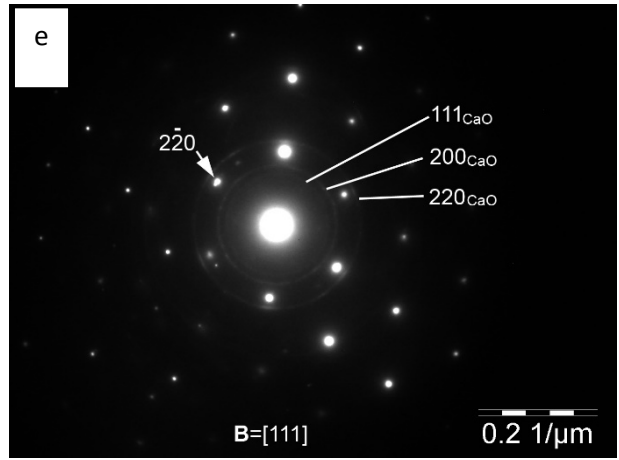

**Figure 1.** TEM image (a) and corresponding SAED pattern (b) of a $CaF_2$ grain close to the [-112] zone axis at the beginning of observation at low magnification (low e-beam dose); TEM image (c) and corresponding SAED pattern (d) of the same $CaF_2$ grain after ca. 60 s of observation under focused e-beam (high e-beam dose); (e) SAED pattern of the same crystal grain in a different orientation (close to the [111] zone axis) after ca. 120 s of observation under focused e-beam (high e-beam dose).

*After irradiation with iodine ions at 23 MeV*

In the case of all the analyzed samples, the SHI irradiation effects could only be revealed for high values of objective lens defocus, either underfocus (df<0) or overfocus (df>0). To be able to correctly interpret the TEM images we note that for the TEM images obtained in the diffraction contrast mode, the features exhibiting bright Fresnel fringe correspond to regions in the sample with lower density than the surrounding matrix (voids). The contrast is opposite when the objective lens is overfocused, the image showing a dark Fresnel fringe on the side of the lower-density material.

The SHI irradiation effects are most visible in the case of the sample irradiated at the highest energy of 23 MeV. When observing the irradiated crystal grains in no-tilt or slightly tilted orientation, the underfocused images exhibit doublets of bright dots of 2-3 nm size (Figure 2). The separation distance between the two bright dots differs across one crystal grain, being smaller close to the grain border and larger for the pairs of dots inside the grain away from the grain border. We attribute these pairs of bright spots to the extremities of the irradiation tracks crossing the crystal grain and intercepting the two limiting surfaces (bottom and upper surface). The variation of the separation distance with the location of the bright dots goes in line with the thickness increase of the analyzed wedge-shaped grains. The SHI tracks consist of heavily disordered material regions which, according to the mentioned Fresnel contrast criterion, correspond to regions of lower density within the matrix (fig. 2).

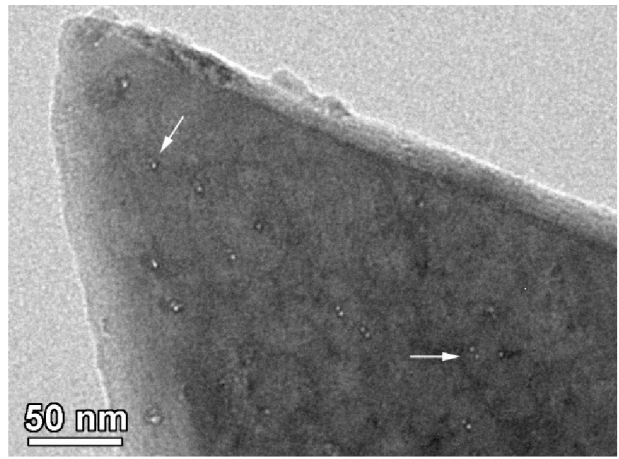

**Figure 2.** Nanometric features revealed as pairs of bright spots exhibited by the $CaF_2$ crystal grains irradiated at 23 MeV.

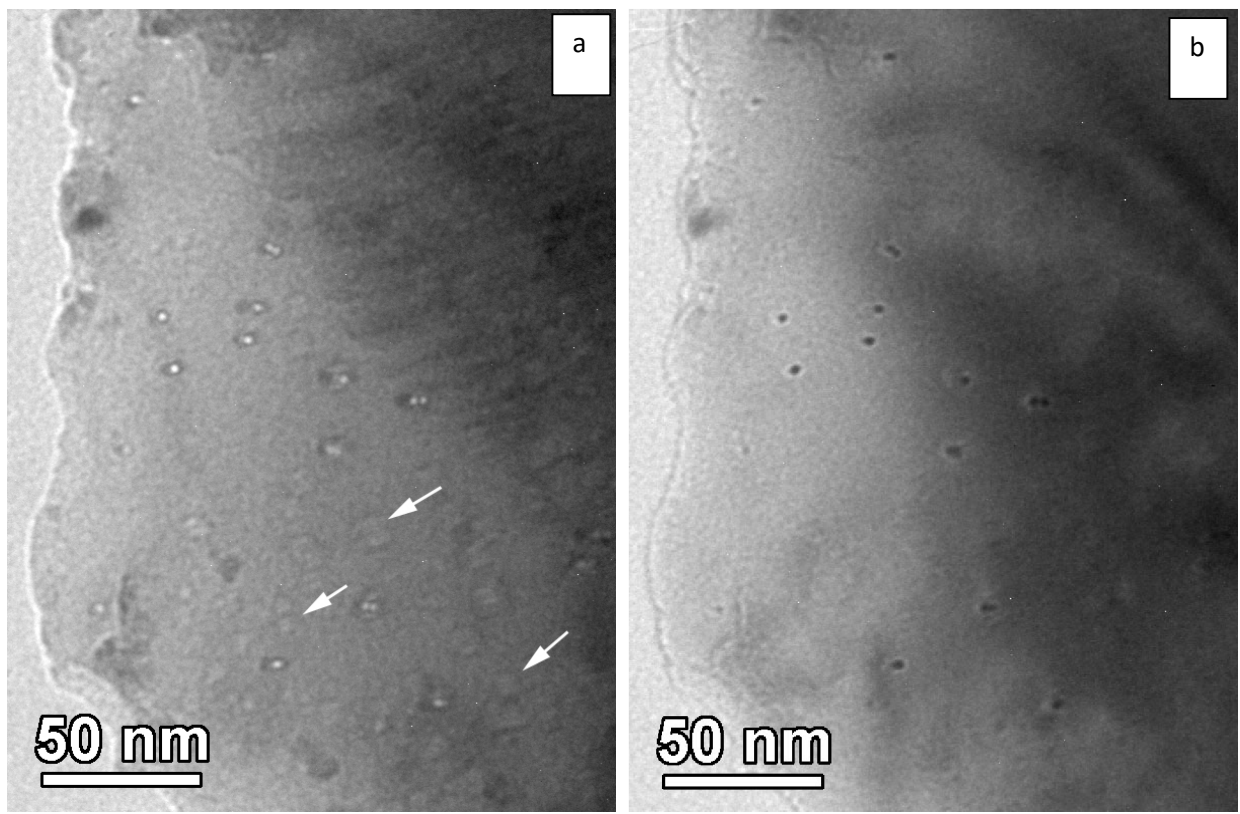

**Figure 3.** Under-focus (a) and over-focus (b) TEM images reveal low density material inside the irradiation tracks represented by the white and, respectively, the black dots. White arrows point to electron beam-induced voids.

In order to get a side view of the irradiation tracks, the sample has been tilted inside the microscope up to 40 degrees with respect to the electron beam. The TEM images obtained in underfocus and overfocus conditions then reveal a certain structuring of the irradiation tracks along their trajectory. They show a contrast modulation with 2-4 maxima along their length.

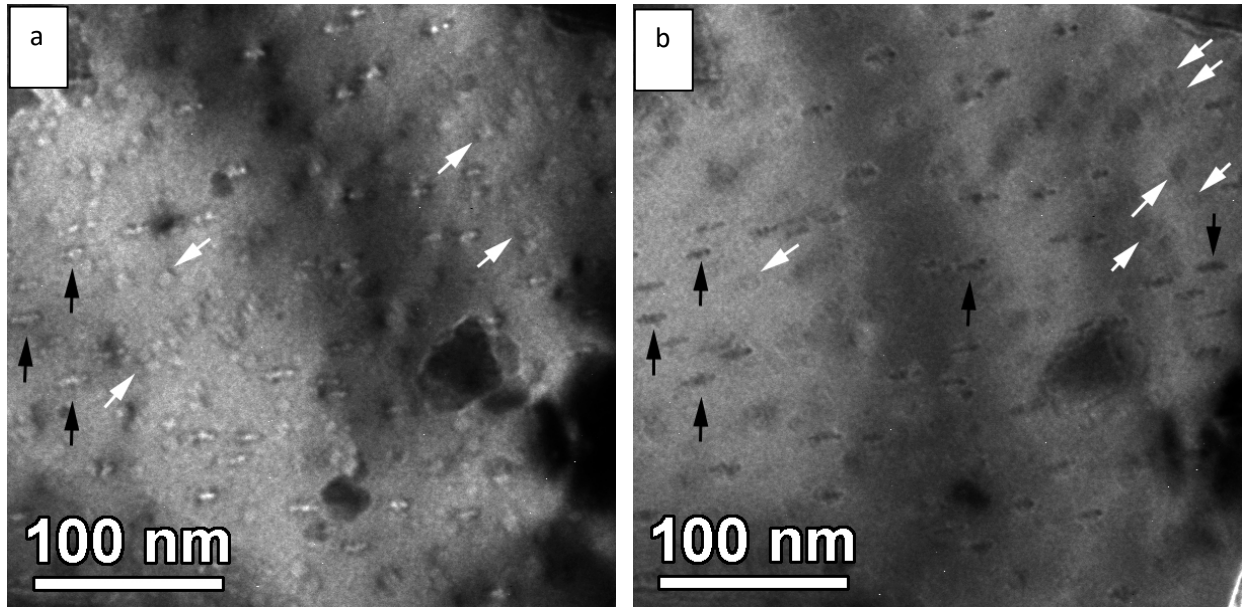

**Figure 4.** Underfocused (a) and overfocused (b) TEM image of a tilted $CaF_2$ grain containing irradiation tracks (black arrows) and an array of electron beam-induced voids (white arrows).

The prolonged observation in the TEM of the SHI irradiated sample leads to the formation of two kinds of defects apart from the irradiation tracks. These defects have already been reported in literature as defects induced by e-beam irradiation during the TEM observation: i. on one hand aggregated defects (voids) sometimes organized in periodic arrays and ii. dislocation loops located in the (111) planes.

The through focus contrast behavior of the first type of defects proves that they are nanovoids of up to 10 nm size. They are indicated by white arrows in Figure 3 and 4 in underfocus and overfocus condition. It is known from the literature [VT79], [THD05] that, these nanovoids tend to organize, with the increase of the irradiation dose and the observation time, into a cubic superlattice oriented parallel to the cubic $CaF_2$ lattice. In our case, due to the low dose observation conditions and the early observation moments when the images have been recorded, the induced nanovoids are not yet organized into a superstructure.

The second type of e-beam induced defects are the dislocation loops bordering planar defects. The calcite grain in Figure 5 is oriented close to the [1-10] zone axis (see the inserted SAED pattern). The image in Figure 5a has been acquired in the first seconds of observation, therefore only the irradiation tracks are visible as doublets of white spots. After a few minutes of observation, pairs of planar defects bordered by dislocation loops started to appear under the influence of the e-beam. By comparing with the corresponding SAED pattern, one can notice that these planar defects are situated in the {111} crystallographic plane of $CaF_2$. Such defects have been also mentioned in literature [EA07].

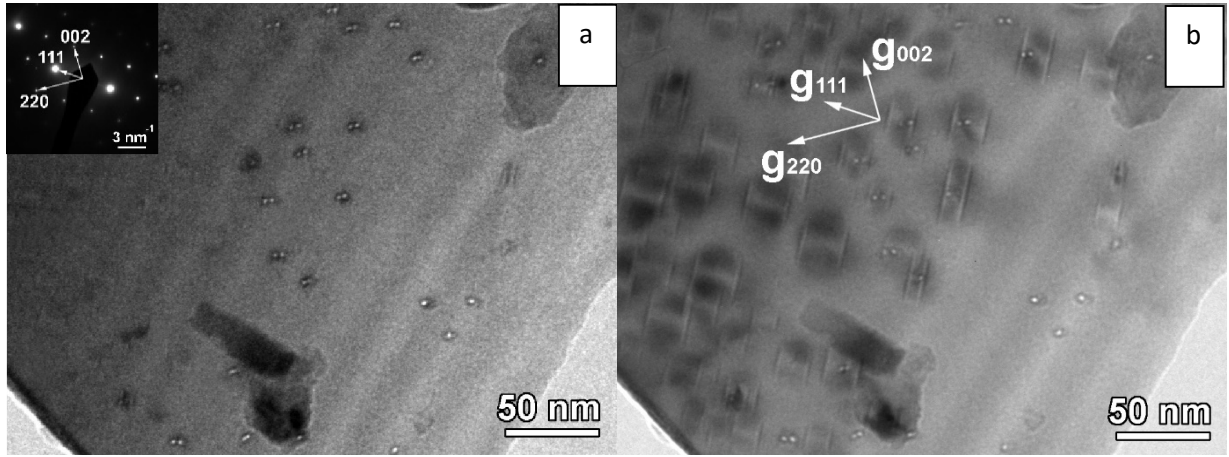

**Figure 5.** (a) Wedge-shaped $CaF_2$ grain oriented close to the [1-10] zone axis (corresponding SAED pattern inserted) containing irradiation tracks imaged as doublets of white dots; (b) e-beam induced planar defects grouped in pairs, bordered by dislocation loops (seen edge-on).

*After irradiation with iodine ions at 15 MeV*

Similar SHI track features have been observed on the samples irradiated at 15 MeV. The bright scattered dots in Figure 6a represent end-on irradiation tracks with a similar size (2-3 nm) as in the case of the sample irradiated at 23 MeV. After a few tens of seconds of observation, e-beam induced defects can be noticed as faceted nanovoids of up to 7-8 nm size (white arrows in Figure 6b).

The side view of the irradiation tracks (Figure 7) reveals the irradiation tracks as having a fine structure of 3-4 aligned nanovoids of 3-4 nm size. E-beam induced faceted nanovoids of up to 8 nm size can also be observed after a few tens of seconds of observation.

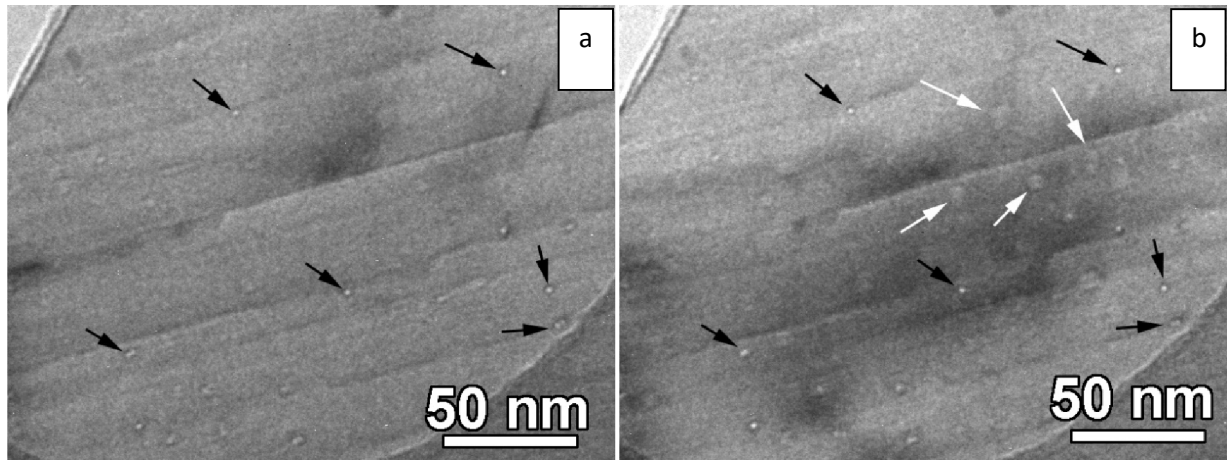

**Figure 6.** (a) Under-focused TEM image showing end-on irradiation tracks revealed as white dots (black arrows); (b) nanovoids induced by e-beam irradiation during the observation in the TEM (white arrows).

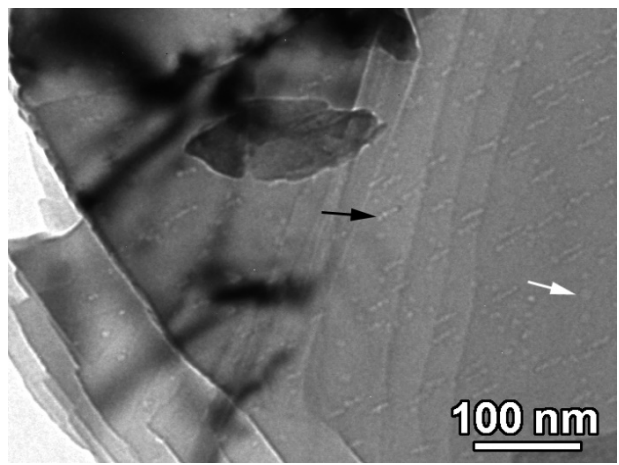

**Figure 7.** Side view of the irradiation tracks (black arrow) in a grain which has been tilted with respect to the electron beam in the microscope. The white arrow points to the regular *in situ* e-beam induced nanovoids.

*After irradiation with iodine ions at 10 MeV*

Similar SHI induced track effects have been observed on the samples irradiated at 10 MeV. Groups of up to 5 bright dots aligned parallel to the vertical direction in Figure 8a represent irradiation tracks in side view orientation, consisting in nanovoids of 5-6 nm size. After a few tens of seconds of observation, e-beam induced defects can be noticed as faceted nanovoids of up to 8-9 nm size (white arrows in Figure 8b).

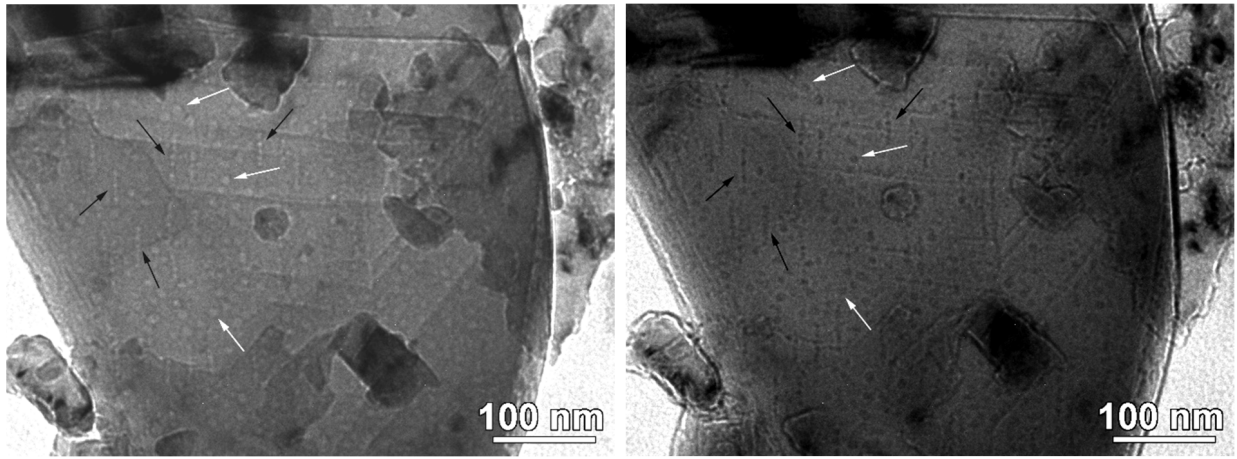

**Figure 8.** Under-focused (a) and overfocused (b) TEM images showing irradiation tracks revealed as groups of 4-5 white dots (black dots when overfocused) aligned parallel to the vertical direction (pointed by black arrows); the white arrows indicate nanovoids induced by e-beam irradiation during the observation in the TEM.

### 3.2. In-situ TEM results

On a careful examination of the ion tracks in the sample irradiated at low energy (10 MeV I), we have observed that, apart from the already mentioned defects induced by the e-beam, there is a certain morphological evolution of the ion tracks themselves under the electron beam. We have noticed that, although initially not visible, the irradiation tracks are only revealed after a few tens of seconds of observation under the e-beam, inside grains which apparently contained no ion tracks. The images in Figure 9 are taken from the same $CaF_2$ grain at different moments starting from the beginning of the observation in the microscope. The image recorded during the very first seconds of observation in underfocus condition does not exhibit any visible ion track. Without changing the illumination conditions or defocus of the objective lens, the ion tracks started to become visible after about 30 seconds of observation, while after 2 minutes they show quite well defined shapes.

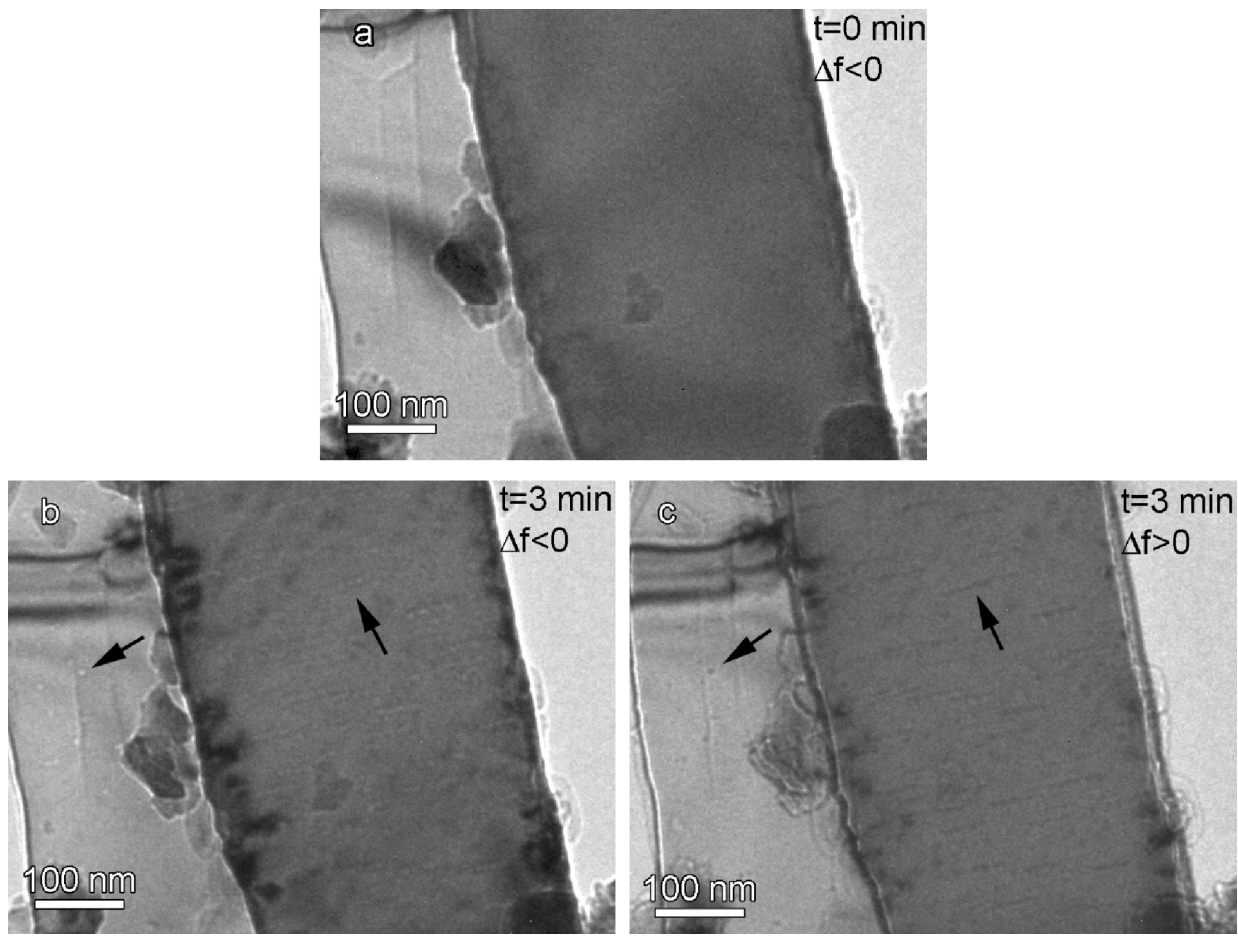

**Figure 9.** TEM image of a CaF$_2$ grains irradiated at 10 MeV: (a) at the beginning of the TEM observation, underfocused image; (b) after 3 minutes of TEM observation, underfocused image; after 3 minutes of TEM observation, overfocused image.

In Figure 9b we present the TEM image after 3 minutes of observation, at the same defocus condition as in Figure 9a (recorded directly after imaging started). The irradiation tracks are now clearly visible, showing the morphological features described previously, both in underfocused and overfocused conditions (Figures 9b and c, respectively).

We mention that the e-beam dose during the TEM observation was maintained at low values and that the typically observed e-beam induced defects in CaF$_2$ have not yet been formed before the ion tracks became visible. This experiment proves that although the ion tracks are produced even for low energy values (10 MeV) as confirmed by the RBS/c measurements (see section 3.4), their spatial extend and, in particular, their contrast is too low to be imaged by TEM in their native state, despite the TEM point resolution which is well below 1 nm for all the commercial TEMs. Any attempt to adjust the TEM observation conditions in order to reveal the sub-nanometric ion tracks results in increasing the e-beam dose which has as direct consequence the rapid modification of the ion tracks from their initial state. From the TEM point of view, the as-produced ion tracks consist of unobservable thin filaments (of the order of 1 nm or below, as revealed by RBS/c) of disordered matter inside the $CaF_2$ lattice. These are only revealed in TEM after a few seconds of e-beam irradiation which induces a process of atomic relaxation and rearrangement along the tracks into rows of nanovoids. Accurately measuring the size of the ion tracks by TEM therefore becomes a delicate issue, since they continuously evolve under the e-beam, their size reaching a saturation value which may be higher than the one measured by RBS/c.

Moreover, as these fine morpho-structural features (ion tracks) are only visible under strong defocus conditions with the help of the bright or dark Fresnel fringe, it is important to mention that the measured size on the TEM micrographs depends on the amount of defocus. In order to illustrate this, in Figure 10 we present a through-focus series of TEM images from the same area of a $CaF_2$ containing ion tracks. Two different values of underfocus (Figures 10a and b) and overfocus (Figures 10c and d) have been used to show that the larger the defocus the larger the measured size of the ion tracks. For the two ion tracks marked with arrows in the central part and bottom part of the micrographs, the measured sizes are 4.1 nm in Figure 10a (small underfocus), 6.6 nm and 5.8 nm in Figure 10b (large underfocus), 7.3 nm and 5.8 nm in Figure 10c (small overfocus), 8.2 nm and 7.4 nm in Figure 10d (large overfocus). From the spread of the data and the fact that the ion tracks are modified by the e-beam illumination, it results that one cannot obtain reliable information on the tracks size from the TEM investigations with an accuracy better than 2-3 nm, which is of the same order of magnitude as the ion tracks. This might explain the contradictory quantitative results obtained by different techniques and authors reported in literature. Although the quantitative information obtained by STEM imaging of the ion tracks is more reliable (section 3.3.) since in this case we deal with the mass-thickness contrast instead of the Fresnel fringes, neither this observation mode prevents morphological transformation of the ion tracks under the e-beam.

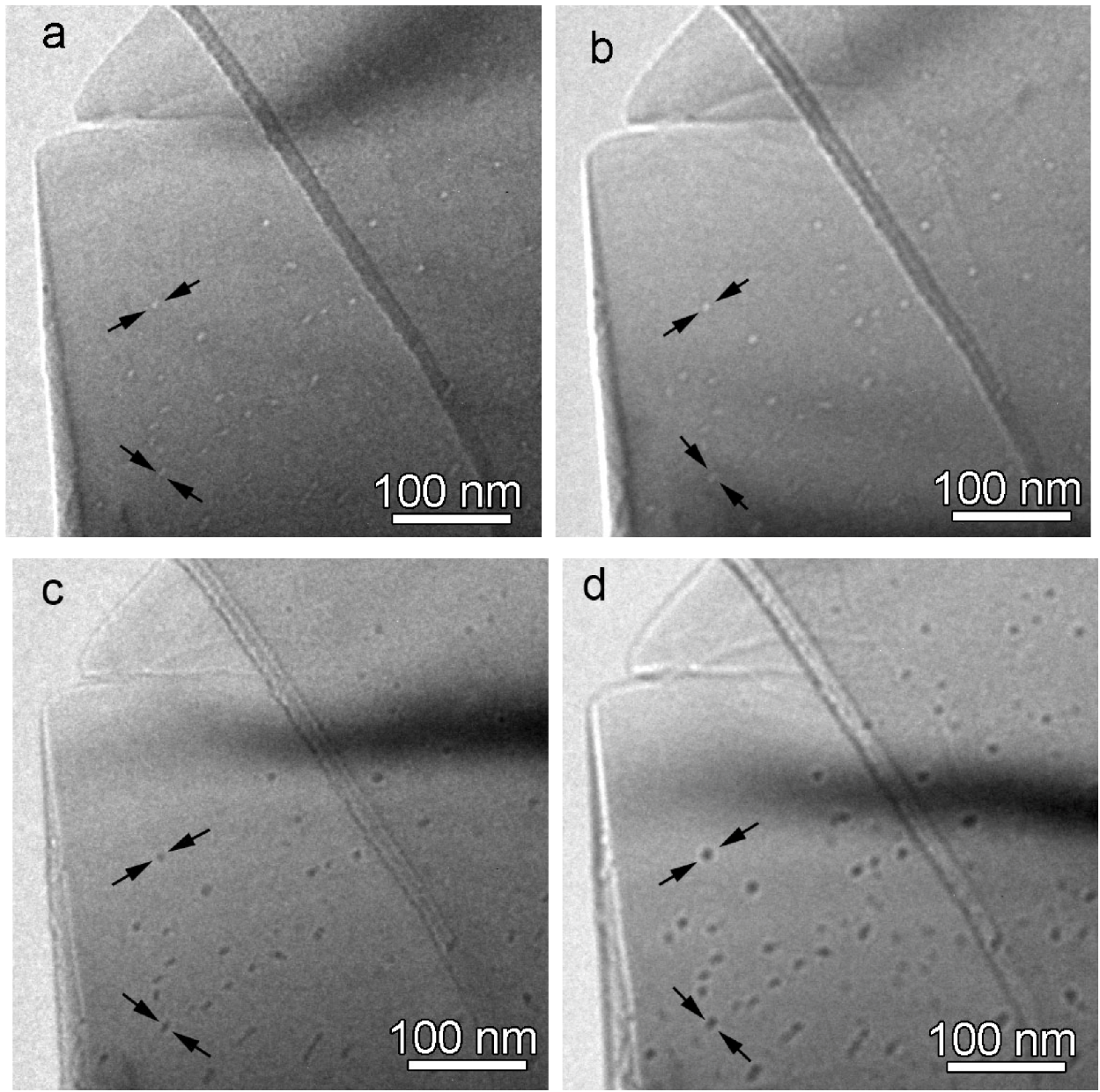

**Figure 10.** Through-focus series of TEM images from the same area inside a $CaF_2$ grain containing ion tracks: (a) $\Delta f_1 < 0$; (b) $\Delta f_2 < 0$, $|\Delta f_1| < |\Delta f_2|$; (c) $\Delta f_3 > 0$; (d) $\Delta f_4 > 0$, $\Delta f_4 > \Delta f_3$.

## 3.3. STEM-EELS results

During the previous TEM investigations performed in low e-beam conditions using the JEOL 200 CX TEM provided with a tungsten filament, we have clearly identified the specific irradiation defects in the $CaF_2$ grains. Therefore, we are able to discriminate the defects induced by SHI irradiation, ion tracks as rows of nanovoids, and those induced in situ, by e-beam irradiation during the TEM observation, undecidedly reported in literature either as voids or Ca colloids, or more generally, aggregated defects.

The STEM working mode is based on the mass-thickness contrast, meaning that signal intensity is proportional to the local $<Z>^2$ and the local sample thickness. On the other hand, EELS is an analytical technique providing qualitative and quantitative information regarding the chemical composition of the sample. By coupling STEM and EELS modes in the Spectrum Imaging (SI) mode, one can get data packages containing both morphological and spectroscopic information. In STEM-EELS-SI, the focused electron beam is scanned across the selected area while an EEL spectrum and morphologic information are acquired in each pixel. EEL spectra and composition maps can be afterwards extracted from the SI data cube with a space resolution which, in certain conditions, can go down to atomic resolution. We have applied the STEM-EELS analysis in order to get coupled chemical and morphological information on the irradiation defects induced by the heavy ions and the electron beam. The analyzed grains contain both SHI and e-beam induced defects. The STEM images have been acquired using the annular dark field (ADF) detector with a collection angle of 180-730 mrad.

On the STEM-ADF image in Figure 11, the e-beam induced defects and the 23 MeV I ion tracks are indicated with white arrows and black arrow, respectively. For both kinds of defects, the characteristic contrast is darker than the surrounding area, indicating a local deficit of matter in both cases, which is in agreement with the conclusions derived from diffraction contrast of the underfocused/overfocused TEM images. The size of the observed defects, as measured on the STEM image, is in the range 2-4 nm for the irradiation track and about 10-15 nm or more for the e-beam induced defects (defect average size proportional to the time spent below the e-beam), confirming the values resulted from the TEM study. Typical intensity line profiles across the ion tracks are presented correspondingly below the STEM images in Figure 11.

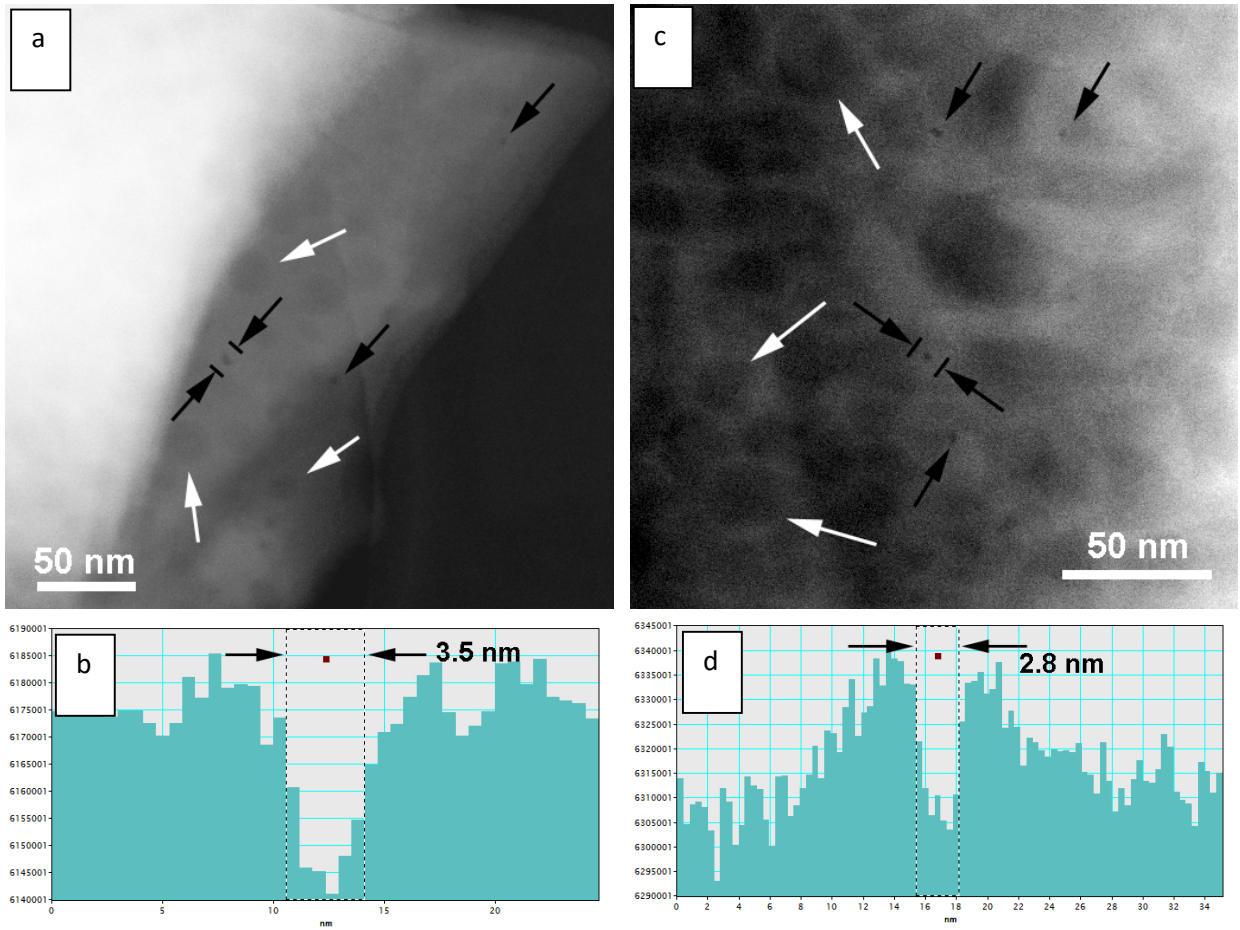

**Figure 11.** STEM ADF images at different magnifications (800 kx and 1200 kx) of a $CaF_2$ grains containing irradiation tracks (black arrows) and e-beam induced defects (white arrows). Intensity line profiles across the irradiation tracks indicate their size (measured as the FWHM of the profile) in the range of 2-4 nm.

The possible chemical modifications associated with the irradiation-induced defects and the spatial distribution of the identified chemical elements have been investigated by EELS in STEM mode (Spectrum Imaging). On the STEM-ADF image in Figure 12a (the same area as the one in Figure 11a), the green rectangle marks a region containing one typical ion track (pointed by the black arrow) as well as defects created in situ by electron beam irradiation (pointed by white arrows). As previously mentioned, these defects show a darker contrast with respect to the surroundings, indicating a deficit of matter (void).

The total EEL spectrum extracted from the whole SI area reveals the chemical elements Ca and F present in the sample. After removing the background, the local intensity of the Ca and F signals is mapped out as Ca and F elemental maps presented in Figure 12b. The two elemental maps have been also overlapped into a composed colored map to check for local chemical segregation. We have used the same arrows to point to the areas corresponding to the morphological details indicated in Figure 11a. Two interesting facts can be noticed by correlating the elemental maps with the STEM image:

i. A local increase of the F signal and a deficit of the Ca signal correspond to the ion track;

ii. A local deficit of the F signal and an increase of the Ca signal correspond to the e-beam induced defects.

On the composed colored map one can clearly observe the spatial complementarities between the two elemental maps and the compositional differences between the ion tracks and the e-beam induced defects.

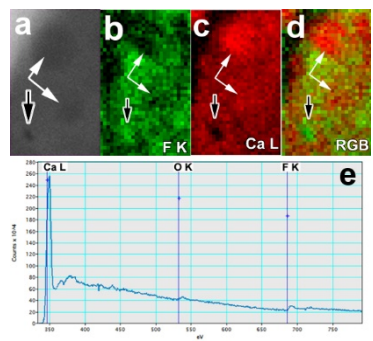

**Figure 12.** (a) STEM-ADF image of a CaF$_2$ grain containing ion tracks and e-beam induced defects. The green rectangle indicates the area from which the Spectrum Image has been acquired. (b) Elemental maps inside the green rectangle illustrating the local content of F and Ca; the composed RGB image has been obtained by superposing the two elemental maps (c) EEL Spectrum extracted from the entire area inside the green rectangle showing the absorption edges of Ca (L$_{2,3}$ edge at 346 eV), F (K edge at 685 eV) and O (K edge at 532 eV).

In addition to the expected Ca and F signals, the O absorption edge at 532 eV can also be noticed in the extracted EEL spectrum. The presence of the O signal is not surprising, since by TEM investigations we have evidenced the morphological and structural modifications of CaF$_2$ under the e-beam and the in situ formation of cubic CaO during the TEM observation of the CaF$_2$ grains (see the SAED patterns of the non-irradiated CaF$_2$ single crystal grains after ca. 1 min. of observation under focused beam, Figure 1e).

Finally, by corroborating the morphological and spectroscopic information resulting from the diffraction contrast images in TEM, mass-thickness contrast images in STEM, STEM-EELS elemental maps we come to the following conclusions:

i. The observed ion irradiation tracks consist of rows of nano-voids (lower Ca signal in the EELS elemental maps) where F may be trapped in gaseous state (higher F signal in the EELS elemental maps).

ii. The e-beam induced defects consist in nanometric volumes where $CaF_2$ dissociates with the likely loss of gaseous $F_2$ and formation of Ca-rich nanometric pockets where Ca is partially oxidized (due to the residual oxygen atoms in the TEM column vacuum). As mentioned also in Ref. [JJ98b], the mass density of Ca (1.54 g/cm$^3$) is lower than the one of $CaF_2$ (3.18 g/cm$^3$), which explains the void-like behavior of the diffraction contrast in the TEM images of the e-beam induced defects. The remaining Ca-rich regions are partially oxidized as proved by the SAED patterns and EEL spectra.

## 3.4. RBS/c results

For each sample, the area irradiated by swift iodine ions was defined by a collimator 3×2 mm$^2$ in size, and the alignment was done on the unirradiated part of the sample. No sample preparation was done, and the ion current was kept low around 1 nA to obtain reliable RBS/c spectra. As shown in fig. 13, it was verified that the RBS/c beam does not introduce defects after prolonged exposure, and that the RBS/c spectra from irradiated samples (containing disorder) can be reliably acquired even after multiple exposures.

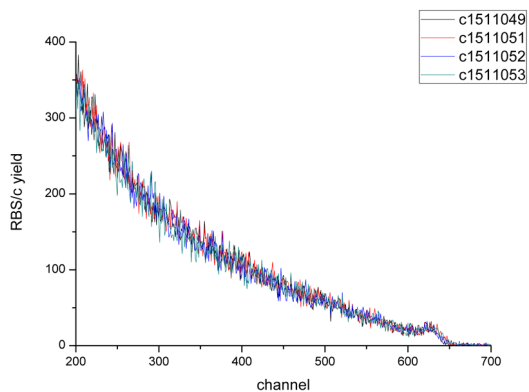

**Figure 13.** Four successive RBS/c spectra obtained from an irradiated CaF$_2$ sample (23 MeV I, fluence 3×10$^{12}$ cm$^{-2}$).

For each iodine energy used, we have irradiated three samples with fluences of 3×10$^{12}$ cm$^{-2}$, 1×10$^{13}$ cm$^{-2}$ and 3×10$^{13}$ cm$^{-2}$, respectively. In total, 9 samples were analysed by RBS/c. After the alignment procedure was accomplished on the unirradiated part of the sample, by a quick angular scan it was verified that the channelling axis is the same on the irradiated part of the sample. For each channelling RBS/c spectrum, a random one was also recorded for the same exposure time. In this way all aligned spectra could be normalized to the random spectra, ensuring all aligned spectra are normalized to the same analysing beam fluence.

In figure 14 we show RBS/c spectra for three different iodine energies, as a function of the applied fluence. It is evident that for each iodine energy used, disorder is introduced into CaF$_2$. Furthermore, by increasing the energy of the iodine ions, disorder is building up faster and eventually saturation is reached at the highest fluence. At that instance, obviously a second stage of damage evolves leading to a pronounced growth of the surface peak (marked by the arrow in fig. 14). For clarity, in fig. 15 we show RBS/c spectra for each fluence, demonstrating again that the observed disordering increases with rise in the kinetic energy of the iodine ions.

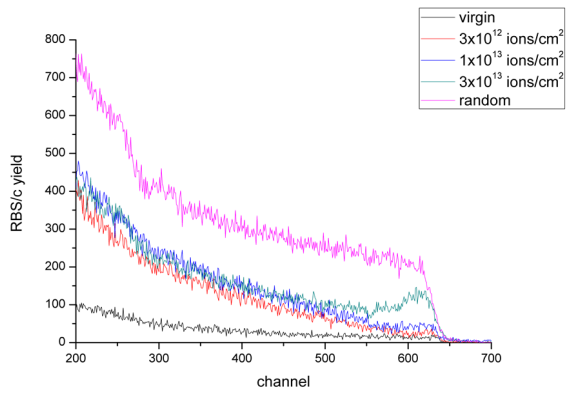

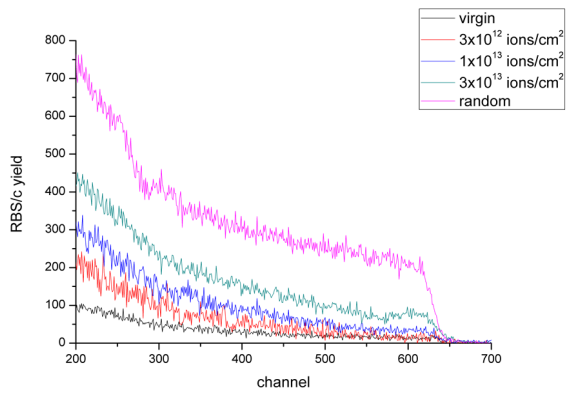

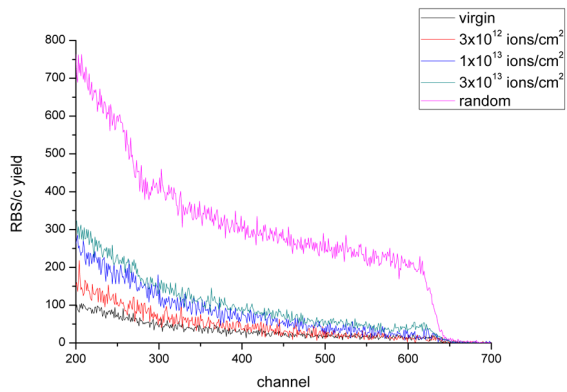

**Figure 14.** RBS/c spectra after a) 23 MeV I b) 15 MeV I and c) 10 MeV I irradiation. The applied fluences were $3\times10^{12}$ ions/cm$^2$, $1\times10^{13}$ ions/cm$^2$, and $3\times10^{13}$ ions/cm$^2$. For comparison, RBS/c spectra from unirradiated sample are given, both in channelling and in random orientation.

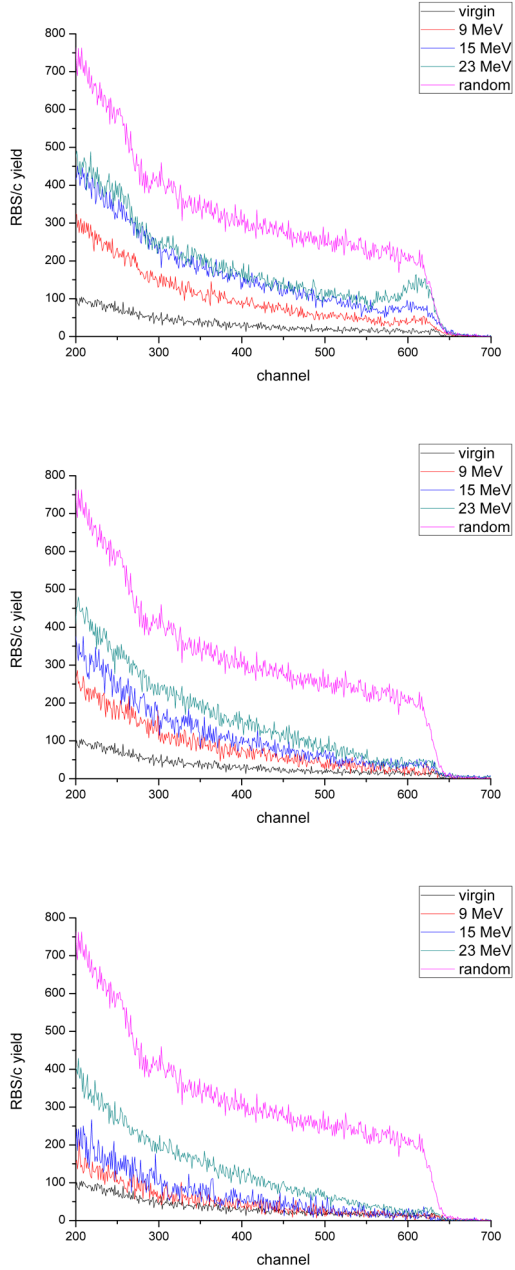

**Figure 15.** RBS/c spectra as a function of ion energy, for a fixed applied fluence of a) $3\times10^{13}$ cm$^{-2}$, b) $1\times10^{13}$ cm$^{-2}$ and c) $3\times10^{12}$ cm$^{-2}$.

Analysis of this kind of RBS/c spectra is usually done in the so-called surface approximation. By neglecting the contribution from the surface peak, the amount of disorder $F_d$ is measured by comparing the backscattering yield from defects ($\chi_i$) relative to the backscattering yield in the random ($\chi_r$) and aligned virgin ($\chi_v$) RBS/c spectra at the position of the surface:

$$F_d = \frac{\chi_i - \chi_v}{\chi_r - \chi_v} \tag{1}$$

Assuming a cylindrical geometry for the ion tracks, it is possible to evaluate the radius of the individual ion tracks by monitoring disorder build-up as a function of the applied ion fluence, and fitting the resulting data with a Poisson formula (thus taking into account ion track overlap):

$$F_d = \alpha \left(1 - e^{-R^2 \pi \Phi}\right) \tag{2}$$

where $\alpha$ is a normalization factor.

It can be immediately realized this approach is not adequate for the present set of experimental data. While it is clear that the damage is correlated to the electronic stopping power of the respective projectile, and the fact that ion tracks are observed by TEM for all used iodine ion energies, direct backscattering on defects within Ca edge is negligible. However, damage observed by RBS/c due to dechannelling of the probing beam indicates damage within the F sublattice because direct backscattering of F atoms contributes to the RBS/c spectrum only below the F edge. In order to verify this, we have performed an RBS/c analysis using 1 MeV protons on the sample irradiated with 23 MeV I at a fluence of $3\times10^{12}$ ions/cm$^2$. Due to different mass and energy of the probing ion beam, an enhanced sensitivity to backscattering from displaced fluorine atoms is achieved, as shown in Fig. 16. In this case, it is evident that both dechannelling and direct backscattering from the disordered fluorine lattice occur, and in principle, the standard approach (i.e. surface approximation) could be applied to the fluorine edge in the RBS/c spectra. The pronounced step at the position of the fluorine surface indicates that most of the fluorine sublattice is disordered, in agreement with results from 2 MeV He RBS/c.

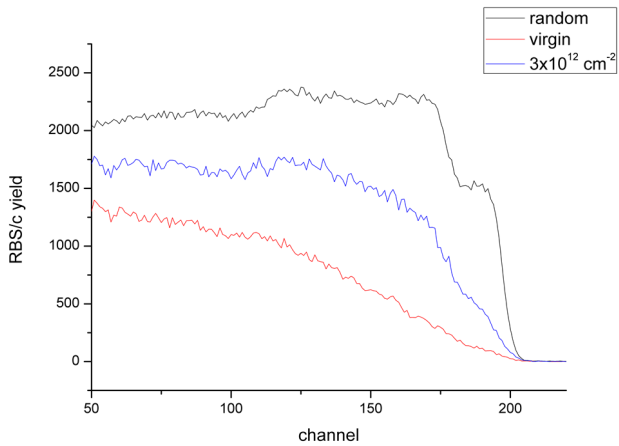

**Figure 16.** RBS/c spectra of samples irradiated with 23 MeV I ions at a fluence of $3\times10^{12}$ ions/cm$^2$, obtained by 1 MeV proton probing beam.

Since dechannelling also depends on the defect concentration [nastasi], dechannelling below Ca edge as a function of applied iodine fluence can be used for ion track measurement. Dechannelling here we define as the slope of the aligned 2 MeV Li RBS/c spectra between channels 400 and 600. Applying Poisson formula (Fig. 17), we estimate the ion track (i.e. disorder within fluorine sublattice) radii to be: r = 3.5 ± 0.1 nm for 23 MeV I, r = 1.6 ± 0.2 nm for 15 MeV I and r = 1 ± 0.2 nm for 10 MeV I irradiation. For the 10 MeV I tracks, saturation of disorder below 1 cannot be ruled out ($\alpha < 1$), but higher iodine fluences would be needed to establish this with certainty.

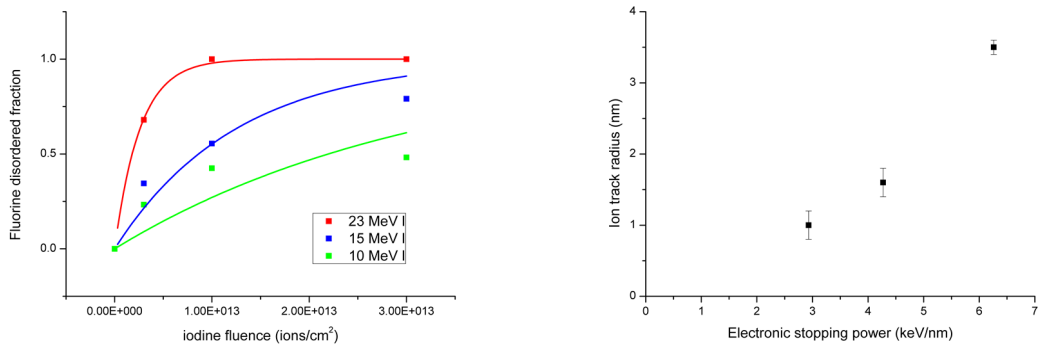

**Figure 17.** (a) Disordered fraction within the fluorine sublattice evaluated from the dechannelling RBS/c yield as a function of the applied iodine fluence for 23 MeV I (red), 15 MeV I (blue) and 10 MeV I (green) irradiation. (b) Ion track size increase with electronic stopping power.

### 3.5. AFM and ToF-ERDA results

Surface ion tracks produced by grazing incidence 23 MeV iodine ion irradiation shown in fig. 18a exhibit the typical, well defined nanohillock-like structure aligned along the SHI trajectory. This is in line with previously published data on surface ion tracks in CaF$_2$ after grazing incidence with 103 MeV Pb ions [SA08]. The observed ion-track density at a fluence of 15 ion impacts per μm$^2$ indicates a surface track formation efficiency close to one.

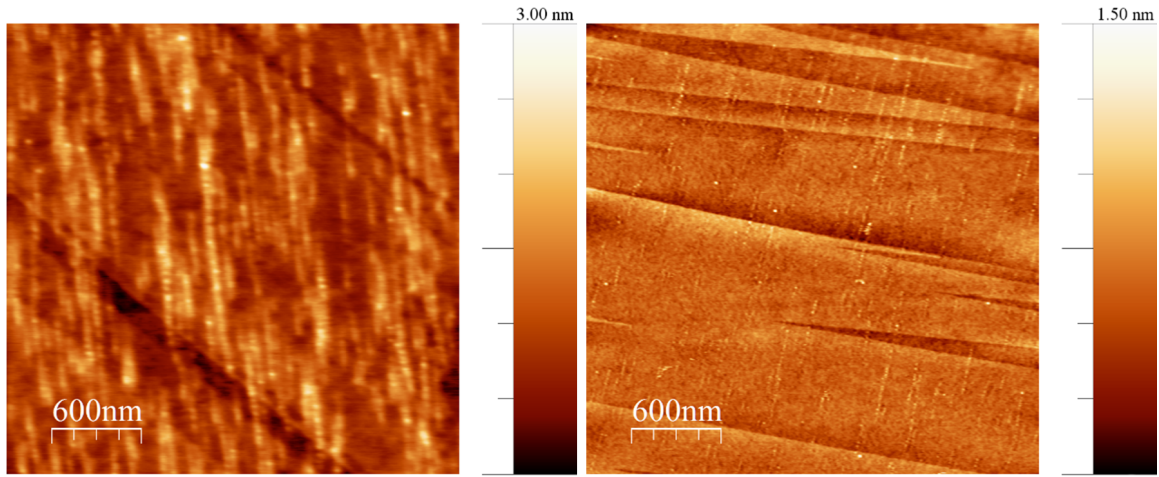

**Figure 18.** AFM images of the CaF$_2$ (111) surface after irradiation with (a) 23 MeV I at 1° incidence angle and fluence of 15 ion impacts/µm$^2$, (b) 10 MeV I at 1° incidence angle and fluence of 10 ion impacts/µm$^2$.

After 10 MeV I irradiation under the same grazing incidence angle of 1°, the resulting surface ion tracks show the same internal structure but they appear much fainter, as can be seen from Fig. 18b. Images have been acquired by the same AFM tip and on the same day, but we cannot rule out changes in the AFM tip quality during the measurements (fig. 18a has been acquired before fig. 18b). The main point is however that also here, the observed density of the surface tracks is in agreement with the applied fluence of 10 ion impacts per µm$^2$ indicating that the track production efficiency is still close to one. Because one would expect reduced efficiency close to the track formation threshold, we conclude that electronic stopping power of 2.9 keV/nm (corresponding to 10 MeV I irradiation) is still well above the threshold for surface ion track formation. This is in agreement with the observation in ref. [YYW14], although thresholds for surface ion tracks produced in the normal and under the grazing geometry probably cannot be compared directly [MK15].

As shown previously, the formation of surface ion tracks after grazing incidence irradiation can be accompanied by depletion of one of the material constituents, resulting in non-stoichiometric surface state [MK15], [MK16]. For this reason, in-situ TOF-ERDA was performed using 23 MeV I ions under the 1° grazing incidence geometry. As shown on fig. 19, even after high fluence iodine irradiation, the stoichiometry of the CaF$_2$ surface remains stable. This is not necessarily at variance with the STEM-EELS results presented in section 3.3 because information obtained from the ToF-ERDA is non-local.

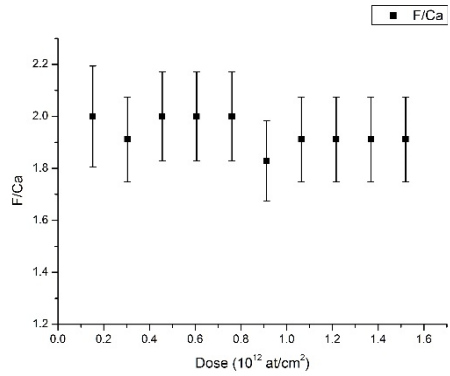

**Figure 19.** Calculated F/Ca ratio from offline analysis of the in situ ToF-ERDA measurements, obtained from the first 10 nm of the $CaF_2$ sample. Measurements were performed using 23 MeV I ions at 1° grazing incidence angle.

## 4. Discussion

### *4.1. Discussion of TEM results*

We begin by comparing TEM results obtained in this study with previously published data [JJ98], [NK05], [SAS07] and related analysis using the ITSM [MT12b], [YYW14] and the ATSM [GS00], [MK12], [GS13], [GS15]. The observation of surface as well as bulk ion track in $CaF_2$ after irradiation with 10 MeV I, yielding an electronic stopping as low as 3 keV/nm, is completely unexpected. From previous TEM studies, the threshold for ion track formation was evaluated to be at 9.5 keV/nm for low velocity irradiation [JJ98], [GS00], while high velocity irradiation data indicated an even higher threshold around 20 keV/nm [NK05], [SAS07]. Such a large value of high velocity threshold, and the small size of the observed ion tracks was dismissed by Szenes as erroneous experimental data [GS13]. While it is true that TEM observation of ion tracks in $CaF_2$ is challenging, due to the sensitivity of the material to e-beam irradiation as demonstrated in the present work, and due to known difficulties in evaluating the ion track diameter [JJ98b], objections of that kind cannot hold when clear experimental evidence for ion tracks is given. Here we note that the ion track sizes obtained in this study (data from section 3.1.) are comparable to other studies, with the exception of experiment with 40 MeV $C^{60}$ cluster ion irradiation, as shown in Fig. 20. The slightly larger ion track diameter for lowest energy irradiation (10 MeV I) is in our opinion not related to the nuclear stopping power. Besides here presented RBS/c data that show monotonic increase of disorder with increasing electronic stopping power, in a previous study with 20.2 MeV Au$^{4+}$ ions [JJ98], data point with nuclear and electronic stopping power of comparable magnitude is indistinguishable from other irradiations with dominant electronic stopping power. Therefore, slightly larger ion track diameters cannot be caused by nuclear stopping power which is one order of magnitude lower than the electronic stopping power in all of our irradiations. Even if we assume that 100% of nuclear stopping power is transferred into the thermal spike, while only 20% of the electronic stopping power (a hypothesis that can be accommodated only within the ATSM), it would still not be possible to justify a track diameter of 5 nm after irradiation with 10 MeV I. At present we cannot explain this slight increase in ion track size after 10 MeV I irradiation, but as shown after *in situ* TEM measurements, these measured ion track radius values are prone to uncertainties related to specific imaging conditions despite our efforts to keep imaging conditions similar for all the samples investigated. These measurements were done under low irradiation conditions, in order to limit or postpone the e-beam damage, and track sizes have been estimated in the saturation conditions. Here we have to reiterate the warning of Szenes [GS13] because TEM results reported in previous works could also be influenced by sensitivity of the $CaF_2$ to e-beam irradiation. For that reason, RBS/c data that show good reproducibility (Fig. 13) might be better for thermal spike analysis than TEM data. Although RBS/c measurements also induce electronic excitations, the repeatability of the measurements indicate stability of the ion tracks during these measurements, probably due to much lower energies of the secondary electrons generated by the probing beam [NAM15]. This way, it could be said that by RBS/c pristine ion tracks could be observed, while during the TEM ion tracks undergo certain development process.

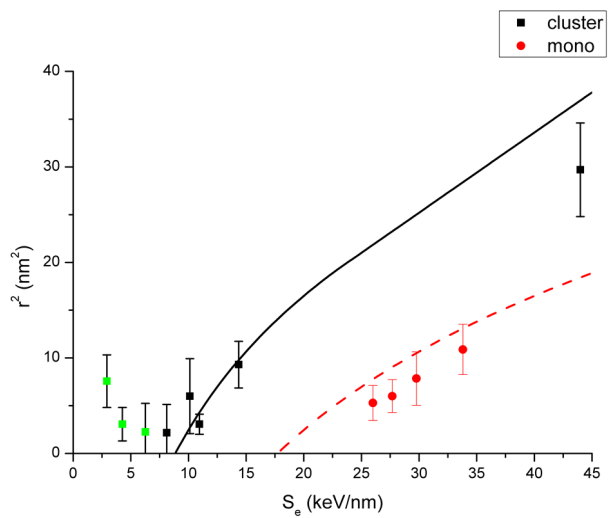

**Figure 20.** Track sizes in CaF$_2$ observed by TEM. New data from the present study (only TEM data from section 3.1.) shown in green. Old data from refs. [JJ98], [NK05] and [SAS07]. Lines are prediction of ATSM [MK12].

Despite the uncertainties with respect to the absolute size determination by TEM, there can be no doubt that well-developed tracks form at stopping powers as low as 3 keV/nm. Clear evidence for this has been presented in this study. Such a low value for ion track formation is not predicted by either one of the thermal spike models. To describe small ion track diameters in CaF$_2$, ITSM invokes boiling as requirement for track formation observable by TEM, with a minimum deposited energy density of 1.7 eV/at [MT12b]. Compared to the melting requirement for nanohillock formation with a minimum deposited energy density of 0.58 eV/at that is achieved at an electronic stopping power of 2.75 keV/nm for very low velocity ion irradiation [YYW14], boiling (hence appearance of ion tracks observable by TEM) should not appear below an electronic stopping power of 8 keV/nm.

The ATSM also faces difficulties in explaining the new experimental data presented here. Previously, the threshold of 9.5 keV/nm for ion track formation determined by TEM studies was attributed either to the absence of the velocity effect [GS00], i.e. the non-existence of a Coulomb explosion contribution at low velocities [GS13b], or to the boiling mechanism like in the ITSM framework [MK12]. In both cases, ion tracks should appear along the ion trajectory at an electronic stopping power of 9.5 keV/nm. Our observation, the appearance of tracks at 3 keV/nm is in clear contrast to these predictions. Even if we take into account that later on the predicted threshold was shifted down to 7 keV/nm (see fig 1 in ref. [GS13]) the large discrepancy remains. Therefore, we can only conclude that the current descriptions of the ion track formation in CaF$_2$ using thermal spike models are not correct.

### *4.2. Discussion of RBS/c results*

We now turn our attention to the RBS/c results. First, in agreement with the TEM data, disordering is observed already after 10 MeV I irradiation and increases with kinetic energy, hence it is an effect due to the electronic stopping power. Disordering is seen as an enhanced RBS/c yield due to dechannelling, while direct backscattering from disordered Ca atoms is observed only for the highest fluences, in the form of an enlarged surface peak. As shown exemplary for 1 MeV proton RBS/c (Fig. 16), this dechannelling originates from the disordered fluorine sublattice, and it should also represent the density of defects [nastasi]. A similar approach was also used recently to assess SHI damage in diamond by simply integrating the RBS/c spectrum within the region of interest [GG15]. By applying the Poisson formula, we have calculated from RBS/c spectra the ion track radius that corresponds to the extent of disorder within the fluorine sublattice after the ion impact (Fig. 17). Qualitatively, the alleged anti-correlation with TEM results is not completely unexpected. Close to the threshold, ion tracks observed by TEM are known to be bigger than tracks observed by RBS/c because of the discontinuous morphology of tracks having diameter below 2 nm [MT06]. Given the difficulties of the TEM analysis revealed in this paper, more than a qualitative agreement cannot be expected at present.

Next, we note that at least for the samples irradiated with 23 MeV I, RBS/c spectra exhibit a saturation behaviour, in which case $CaF_2$ is never fully disordered, in agreement with previous data [MT12b]. This is an indication for a structural recovery of the Ca sublattice during high-fluence irradiation. At the moment it is not known if this is due to a so-called particle assisted prompt anneal (PAPA) mechanism [LTC03] or a swift heavy ion beam induced epitaxial crystallisation (SHIBIEC) mechanism [AB06] operating on already existing ion tracks. The difference between the two mechanisms is that, in the first case, a recrystallization would occur in the wake of the impacting ion (probably the mechanism responsible for the absence of ion tracks in crystalline Si [OO12]), while the second one considers the recrystallization of defects produced by a previous ion impact. Regardless of the origin of this effect, the saturation in RBS/c spectra was also observed before in sapphire [AK08], reproduced here as Fig. 21. Actually, in the same work an enlargement of the aluminium surface peak was also observed, and it was stated that the damage kinetics is a two-step process. The driving force for the enlargement of the surface peak observed in sapphire is most likely due to the suppression of the recrystallization process at the surface and at the crystal-amorphous interface [TA00], [GS05], [AK08].

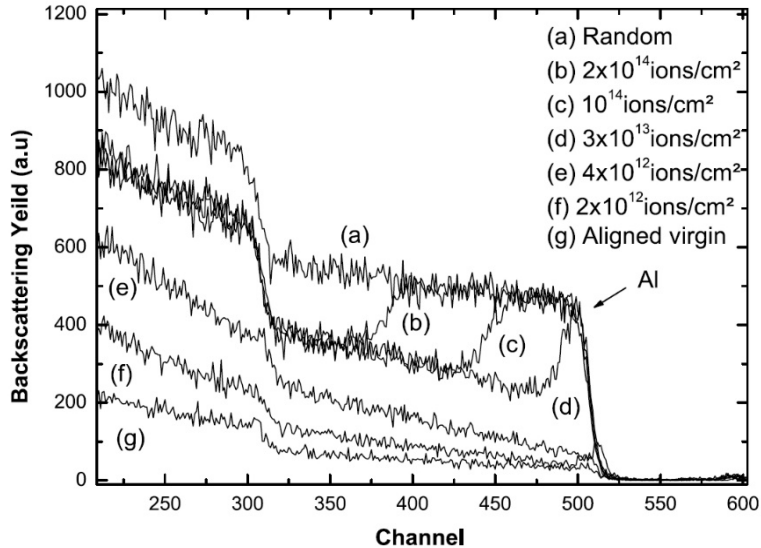

**Figure 21.** RBS/c spectra for sapphire irradiated by 90 MeV Xe. Reproduced with permission from ref. [AK08].

It can be observed (Fig. 21) that up to the onset of the second step, i.e. up to the fluence of $4\times10^{12}$ ions/cm$^2$, enhanced RBS/c yield from the aluminium step due to dechannelling increases with applied fluence. This can be attributed to disordering of the oxygen sublattice (that contributes only to dechannelling within the Al step), and indeed a monotonous increase of the oxygen edge due to backscattering of the RBS/c probing beam can be clearly seen. Saturation of the RBS/c yield is observed for fluences of $3\times10^{13}$ ions/cm$^2$ and above coincides with a fully developed oxygen step, corresponding to a completely disordered oxygen sublattice. Apparently, the onset of the second stage of the damage evolution occurs when the oxygen lattice is fully disordered, but a more detailed analysis would be needed to verify this. Our results indicate that a similar two-stage process is active in CaF$_2$. The reason why it was not observed in the previous study [MT12b] is probably that it occurs only at high fluences.

We note that for all of our RBS/c data points, disordering of the calcium lattice was not observed (except for the surface peak enlargement at very high fluences during the second stage that we do not analyse here). Indeed, as shown by Szenes [GS13], only for stopping power values above 7 keV/nm, disordering of the calcium lattice can be seen. But we also note that for the 2 MeV He RBS/c spectrum of a CaF$_2$ sample irradiated with low fluence high energy ions (375 MeV Pb, dE/dx = 29 keV/nm) shown on Fig. 11a in ref. [MT12b], that dechannelling is substantial while direct backscattering of ions from the Ca lattice is minimal. Only for higher fluences can backscattering from disordered Ca atoms be observed due to disordering of the Ca sublattice. Therefore, we conclude that two separate values of the ion tracks (and corresponding thresholds) can be extracted from the analysis of RBS/c data: one that is related to disordering of the fluorine sublattice and another one that is related to disordering of the calcium sublattice. In our opinion, both values are relevant for the present and previous RBS/c analyses of ion tracks in $CaF_2$ and both cannot be described with a single set of model parameters.

### 4.3. Elemental composition of ion tracks in $CaF_2$

Finally, we turn our attention to the question of the composition of the individual ion tracks in $CaF_2$. Based on previous observations of electron induced defects in $CaF_2$ [EJ83], anion nanovoids were assumed to be the building blocks of intermittent ion tracks in $CaF_2$ observed by TEM [JJ98], [JJ98b], [NK05], [SAS07]. Still, the whereabouts of the fluorine remained an open question. Presumably released as a gas, it was proposed that it is efficiently released from within the ion track [SAS07]. Alternatively, it was proposed by Chadderton [LTC03] that fluorine can remain trapped within the material but outside the ion track.

Our results provide some new insight on this topic. We have shown that by irradiating $CaF_2$ at grazing incidence angle using a 23 MeV I beam, well developed ion tracks can form, as seen by AFM. In situ ToF-ERDA measurements using the same ion beam parameters, provide evidence that the surface composition does not change. Thise evidence rules out the fluorine gas release scenario, and STEM-EELS provides further support that fluorine remains trapped within the ion track. However, the loss of the calcium as seen by STEM-EELS is surprising, and apparently at odds with ToF-ERDA results. But, these results are not necessarily mutually inconsistent, because ToF-ERDA provides information on the macroscopic scale, and STEM EELS on the microscopic. Furthermore, as mentioned in section 4.1, secondary electrons generated by swift heavy ion (in this case also ToF-ERDA beam) have much lower kinetic energies [NAM15] than e-beam used for STEM EELS, and are therefore less damaging to $CaF_2$ structure. In that sense, STEM EELS results provide further insight about development of the ion tracks during the TEM measurements (section 3.2).

We propose following interpretation of our results. Although many of the $CaF_2$ chemical bonds are broken by SHI irradiation, Ca ions remain in place probably due to recrystallization of the Ca sublattice (inferred from RBS/c data [MT12b]), while F ions are disordered, as indicated by RBS/c results reported in the present work. As shown in Fig. 13, the RBS/c probing beam seems to leave unaffected the ion track structure, but in the TEM the ion tracks evolve continuously under the e-beam via radiolysis processes. At the beginning of the TEM observation, no contrast of ion tracks is visible, and this should correspond to the situation that we find by RBS/c measurements, with disordered F and non-disordered Ca sublattices. Only after a few tens of seconds, ion tracks are revealed in the TEM because the broken Ca-F chemical bonds are not restored by the e-beam in the microscope. On the contrary, Ca partially diffuses inside the neighboring $CaF_2$ lattice, and the ion track becomes enriched in F, as indicated by EELS-SI. The Ca out-diffusion from the ion tracks into the surrounding lattice corresponds to the ion track developing process observed by in-situ TEM. The resulting ion track consists then of a channel formed with a lower density of Ca atoms, i.e. enriched in fluorine, which explains the contrast observed by defocus in conventional TEM observation. Recent results on e-beam irradiated $CaF_2$ that indicate a loss of calcium for high enough electron fluences provide further support for such scenario [THD05].

**Conclusion**

By using TEM and RBS/c techniques, we present evidence of ion track formation in $CaF_2$ for electronic stopping power values as low as 3 keV/nm. This suggests melting as requirement for ion track formation since a similar threshold was found for nanohillock formation on the $CaF_2$ surface [YYW14]. Bulk ion tracks observed by TEM after irradiation at the same low stopping power, indicate that current interpretation within the thermal spike mechanism of ion track formation in $CaF_2$ (either related to boiling within the ITSM or an absent Coulomb explosion contribution within the ATSM) is inadequate. Alternative explanations such as the PAPA mechanism as proposed by Chadderton [LTC03] should be checked by future experimental studies.

Hereby we presented a new way of analysing RBS/c data and found clear indications for disordering of the fluorine sublattice at low stopping powers, even under conditions when the calcium sublattice remains undamaged, probably due to recrystallization. The relationship between the previously published RBS/c data [MT12b] and data presented here remains unaddressed, since the former could be re-evaluated using the method demonstrated here. The TEM data presented here shows convincingly that the evaluation of ion track sizes using this technique results in ambiguous results, depending on the imaging conditions. To what extent previous data are subject to these effects remains unknown, but our findings provides support to Szenes questioning of data in that respect.

Surface ion tracks were produced after grazing incidence SHI irradiation for the same low value of electronic stopping power, showing the well-known intermittent structure. Both ToF-ERDA and STEM-EELS experiments provided evidence that fluorine remains within the material when ion tracks are formed in $CaF_2$.

## Acknowledgements

MK, ZS, MJ and SF acknowledge the financial support from the Croatian Science Foundation (pr. No. 8127). Support from the Croatian Centre of Excellence for Advanced Materials and Sensors is also acknowledged. CG and RFN acknowledge the financial support from Core Project PN16-480102. CERIC-ERIC consortium is also acknowledged for the support.